\documentclass[11pt,oneside]{amsart}
\usepackage{geometry}                
\usepackage{graphicx}
\usepackage{amssymb}
\usepackage{epstopdf}
\usepackage{lineno,hyperref}
\usepackage{lipsum}
\usepackage{braket}

\begin{document}

\title{Calculation of the effect of slit size on emittance measurements made by a two-slit scanner}
\author{R. D'Arcy* \\ \emph{UCL, London, UK, WC1E 6BT and Fermilab} \\ \\ A. Shemyakin \\ \emph{Fermilab, Batavia, IL 60510, USA}}
\date{\today}
\thanks{* richard.darcy@desy.de}
\maketitle

\section{Abstract}

Parallel slit-slit devices are commonly used to measure the transverse emittance of a particle beam, selecting a portion of the beam with the front slit and measuring the angular distribution with the rear. This paper calculates the effect of finite slit sizes on measured emittance and Twiss functions in the case of Gaussian spatial and angular distributions of the oncoming beam. A formula for recovering the true emittance from the measured values is derived.

\section{Introduction}

\begin{figure}[h]
    \centering
    \includegraphics[width=100mm]{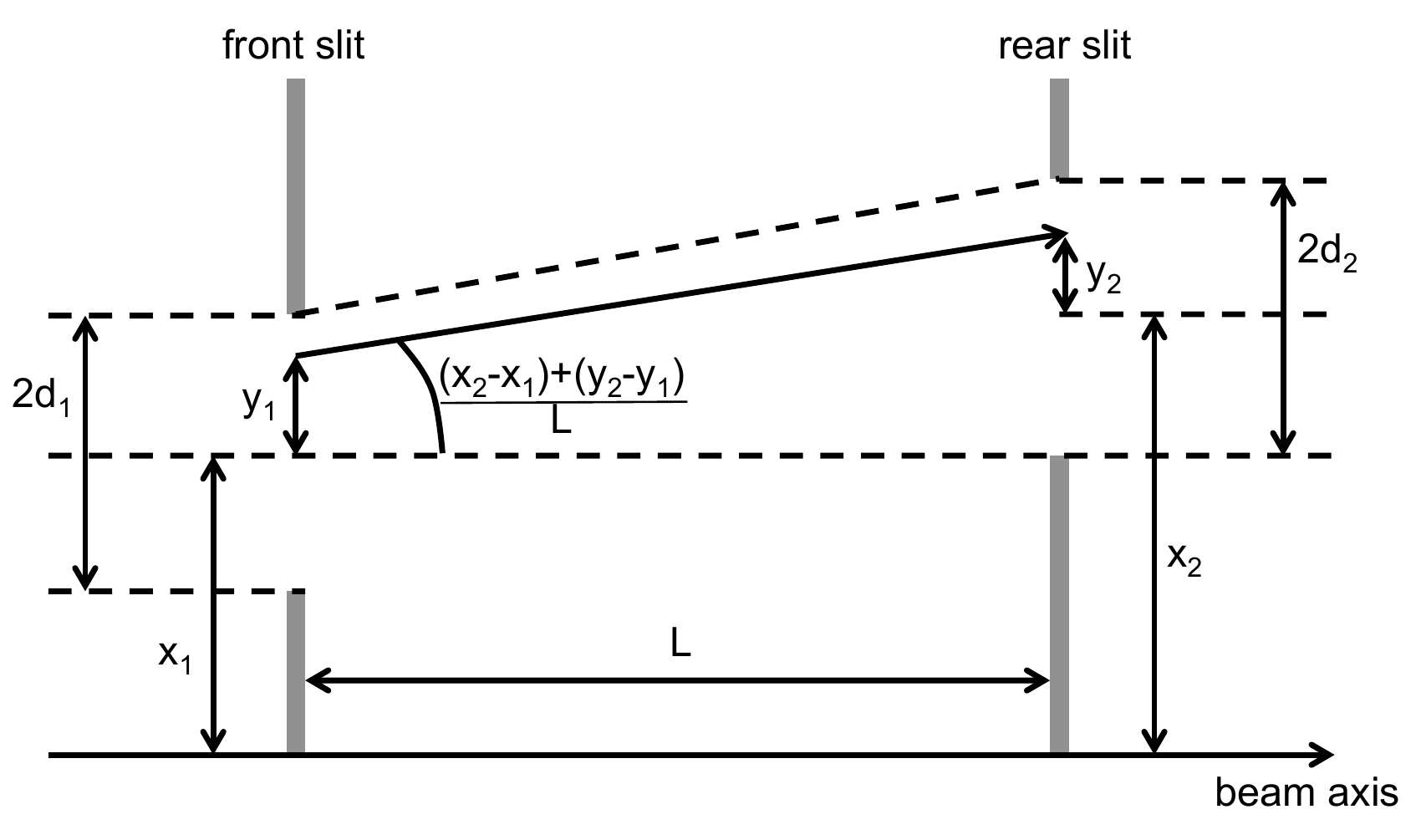}
    \caption{Example geometry of a two-slit collector, displaying notation defined for all calculations in this paper.}
    \label{fig:slits}
\end{figure}

One of many devices used to measure the beam emittance is a two-slit emittance scanner \cite{two-slit}, which consists of two narrow slits separated by a distance $L$ (Fig. \ref{fig:slits}). The beam is sent to the front slit of the scanner, which then cuts out a flat `beamlet'. The transverse beamlet expands proportionally to its initial angular spread and is thus measured by moving the back slit, recording the current passing through both slits with a collector. Repeating the measurement at various front and rear slit positions allows for reconstruction of a full phase space portrait and calculation of the beam emittance.

Alternatively, in an Allison scanner \cite{allison} the beamlet is moved across the back slit, the position of which is fixed with respect to the front slit, by applying a transverse electric field along the beamlet trajectory. If the quality of the electric field is good and effects of secondary particles are minor, both scanners give the same result and are affected by the slit size in an identical manner.

Effect of finite slit size is discussed in \cite{sander}. However, the formula derivation is not presented (only referenced to a private communication), and the formula itself clearly has a typo because dimensions of terms in the sum differ. 

In this paper, we derive formulae for this effect to be directly applied to measurements made with the Allison-type emittance scanner at Fermilab's PXIE LEBT \cite{ibic}.

\section{Assumptions and Notation}
Consider a Gaussian beam distribution with normalised phase density

\begin{equation}\label{eq:dist}
f (x, x') = \frac{1}{2 \pi \epsilon_0} e^{-\frac{x^2 + (\alpha x + \beta x')^2}{2\varepsilon_0\beta}} \quad ,
\end{equation}
\\
\noindent where $\alpha$ and $\beta$ are Twiss functions at the front slit and $\varepsilon_0$ is the real emittance.

The beam phase portrait is measured by a scanner with two infinitely long slits, with width $2d_1$ at the front and $2d_2$ at the rear. The measured emittance may be calculated for such a scanner, assuming that steps of slit motion are much smaller than all relevant dimensions i.e. summing can be replaced by integration.

At each step the position of the slit centres are denoted as $x_1$, $x_2$ and particle coordinates (with respect to the slit centres) as $y_1$, $y_2$, where indices 1 and 2 refer to the front and back slits respectively. 

\section{Calculation}
\begin{figure}[h]
    \centering
    \includegraphics[width=100mm]{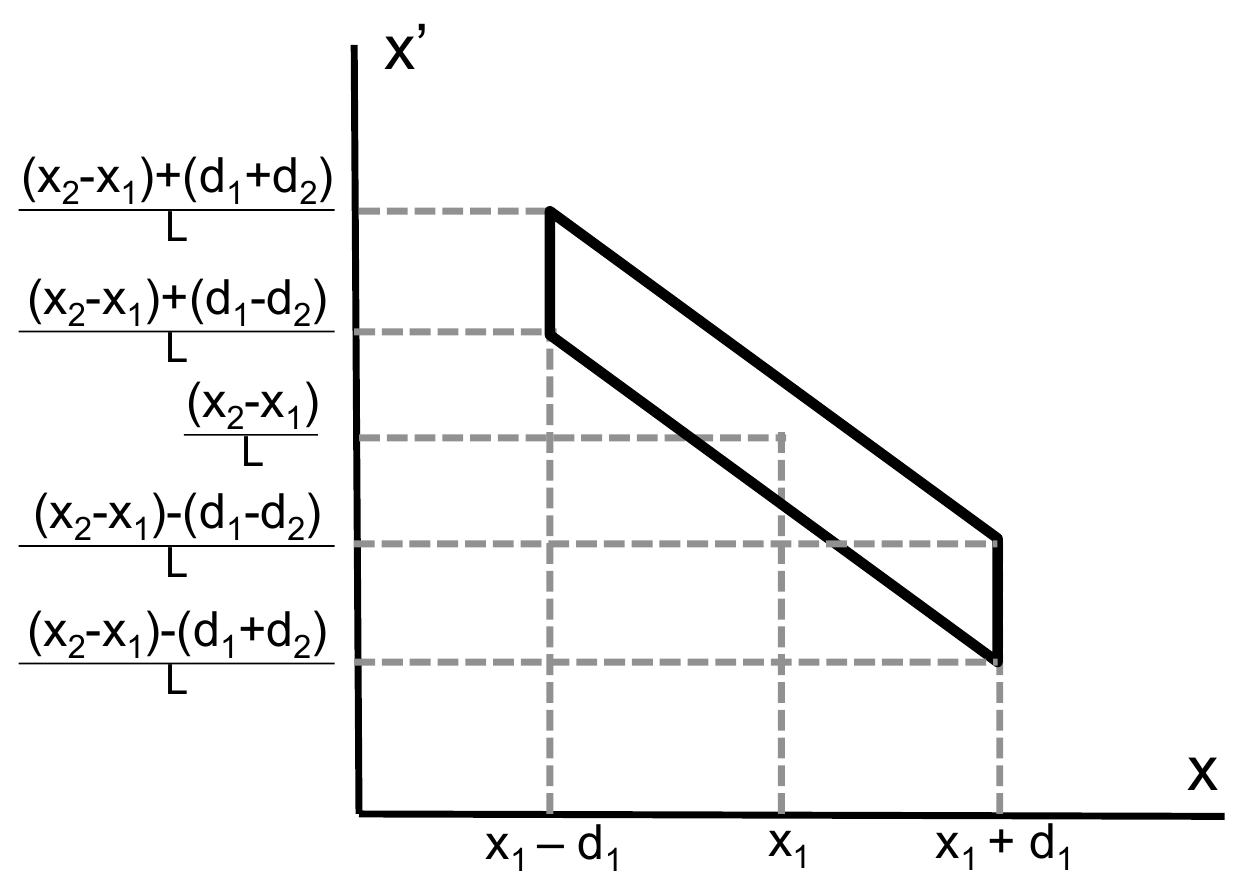}
    \caption{Pictorial representation of the true phase-space acceptance of the front and rear slits, bounded by the given limits.}
    \label{fig:ellipse}
\end{figure}

At a given step of measurement the portion of the beam that reaches the collector $I_c$ is determined by integration of Eq. \ref{eq:dist} over the surface outlined in Fig. \ref{fig:ellipse}.

To simplify calculations, we can integrate over $y_1$ and $y_2$:

\begin{equation}\label{eq:ic}
I_c (x_1, x_2) = \int^{d_1}_{-d_1} dy_1 \int^{d_2}_{-d_2} dy_2 \enspace \frac{1}{L} \enspace f \left (x_1 + y_1, \frac{x_2 + y_2 - x_1 - y_1}{L} \right ) \quad .
\end{equation}
\\

The sum $S_0$, measured with steps $\Delta x_1$ and $\Delta x_2$, can be approximated by the integral

\begin{equation}\label{eq:s0}
S_0 \equiv \left ( \sum \limits_{i,j} I_c (x_{1i},x_{2j}) \right ) \Delta x_1 \Delta x_2 \approx \int^{\infty}_{-\infty} dx_1 \int^{\infty}_{-\infty} dx_2 \enspace I_c (x_1, x_2) \quad .
\end{equation}
\\

This form can be integrated analytically by changing the order of integration, followed by a substitution of variable,

\begin{eqnarray}\label{eq:s0_arr}
S_0 &=& \int^{\infty}_{-\infty} dx_1 \int^{\infty}_{-\infty} \frac{dx_2}{L} \enspace \frac{1}{2\pi\varepsilon_0} \int^{d_1}_{-d_1} dy_1 \int^{d_2}_{-d_2} dy_2 \enspace \frac{1}{L} f \left (x_1 + y_1, \frac{x_2 + y_2 - x_1 - y_1}{L} \right ) \nonumber \\
&=& \frac{1}{2\pi\varepsilon_0 L} \int^{d_1}_{-d_1} dy_1 \int^{d_2}_{-d_2} dy_2 \int^{\infty}_{-\infty} dx_1 \int^{\infty}_{-\infty} dx_2 \enspace e^{-\frac{(x_1+y_1)^2 + (\alpha (x_1+y_1) + \frac{\beta}{L} (x_2 + y_2 - x_1 - y_1))^2}{2\varepsilon_0\beta}} \nonumber \\
&=& \frac{1}{2\pi\varepsilon_0 L} \int^{d_1}_{-d_1} dy_1 \int^{d_2}_{-d_2} dy_2 \int^{\infty}_{-\infty} dx_1 \enspace e^{-\frac{(x_1+y_1)^2}{2\varepsilon_0\beta}} \int^{\infty}_{-\infty} du \enspace \frac{L}{\beta} \enspace e^{-\frac{u^2}{2\varepsilon_0\beta}} \nonumber \\
&=& \frac{1}{2\pi\varepsilon_0 \beta} \int^{d_1}_{-d_1} dy_1 \int^{d_2}_{-d_2} dy_2 \int^{\infty}_{-\infty} dx_1 \enspace e^{-\frac{(x_1+y_1)^2}{2\varepsilon_0\beta}} \int^{\infty}_{-\infty} du \enspace e^{-\frac{u^2}{2\varepsilon_0\beta}}
\end{eqnarray}
\\

The rightmost integral may then be solved using the identity found in Eq. \ref{eq:eu}, leading to

\begin{equation}\label{eq:s0_2}
S_0 = \frac{1}{\sqrt{2\pi\varepsilon_0 \beta}} \int^{d_1}_{-d_1} dy_1 \int^{d_2}_{-d_2} dy_2 \int^{\infty}_{-\infty} dx_1 \enspace e^{-\frac{(x_1+y_1)^2}{2\varepsilon_0\beta}} \quad .
\end{equation}
\\

The same identity is again employed for the integral over $x_1$, followed by definite integrals over $y_1$ and $y_2$. The final result is

\begin{equation}
S_0 = 4d_1d_2 \quad .
\end{equation}
\\

The integrals used to calculate the second moments are

\begin{eqnarray}
S_{xx} &=& \int^{\infty}_{-\infty} dx_1 \int^{\infty}_{-\infty} dx_2 \enspace x^2_1 \enspace I_c (x_1, x_2)  \quad , \\
S_{x'x'} &=& \int^{\infty}_{-\infty} dx_1 \int^{\infty}_{-\infty} dx_2 \enspace \left ( \frac{x_2 - x_1}{L} \right )^2 \enspace I_c (x_1, x_2)  \quad , \\
S_{xx'} &=& \int^{\infty}_{-\infty} dx_1 \int^{\infty}_{-\infty} dx_2 \enspace \left ( \frac{x_2 - x_1}{L} \right ) x_1 \enspace I_c (x_1, x_2)  \quad .
\end{eqnarray}
\\

Proceeding with integration similar to that in Eqs. \ref{eq:s0_arr} and \ref{eq:s0_2}, and utilising the integration identities in Eqs. \ref{eq:ueu} and \ref{eq:u2eu}, the second moments are defined as

\begin{eqnarray}
\Braket{x^2} &=& \frac{S_{xx}}{S_0} = \varepsilon_0 \beta + \frac{d_1^2}{3} \quad , \\
\Braket{x'^2} &=& \frac{S_{x'x'}}{S_0} = \varepsilon_0 \left ( \frac{1+\alpha^2}{\beta} \right ) + \frac{d_1^2 + d_2^2}{3L^2} \quad , \\
\Braket{xx'} &=& \frac{S_{xx'}}{S_0} = - \alpha \varepsilon_0 - \frac{d_1^2}{3L} \quad .
\end{eqnarray}
\\

Finally, the reconstructed emittance and Twiss functions from measured data are given by

\begin{eqnarray}\label{eq:emit_m}
\varepsilon^2_m \equiv \Braket{x^2}\Braket{x'^2} - \Braket{xx'}^2 &=& \varepsilon_0^2 + \varepsilon_0 \left (\beta \frac{d_1^2 + d_2^2}{3L^2} + \frac{1+\alpha^2}{\beta} \frac{d_1^2}{3} - \alpha \frac{2d_1^2}{3L} \right ) + \frac{d_1^2 d_2^2}{9L^2} \nonumber \\
&=& \varepsilon_0^2 + \varepsilon_0 \left ( \frac{\beta}{3L^2} \left [ d_2^2 + d_1^2 \left \{ \left ( 1 - \frac{\alpha L}{\beta} \right )^2 + \frac{L^2}{\beta^2} \right \} \right ] \right ) + \frac{d_1^2 d_2^2}{9L^2} \quad ,
\end{eqnarray}

\begin{eqnarray}
\beta_m &\equiv& \frac{\Braket{x^2}}{\varepsilon_m} = \beta \frac{\varepsilon_0}{\varepsilon_m} + \frac{d_1^2}{3\varepsilon_m} \quad , \label{eq:beta} \\ 
\alpha_m &\equiv& \frac{\Braket{xx'}}{\varepsilon_m} = \alpha \frac{\varepsilon_0}{\varepsilon_m} + \frac{d_1^2}{3L\varepsilon_m} \quad . \label{eq:alpha}
\end{eqnarray}
\\

The relations of Eq. \ref{eq:beta} and \ref{eq:alpha} for the measured Twiss functions may be substituted into that of the measured emittance (Eq. \ref{eq:emit_m}). This conveniently gives the true emittance exclusively in terms of measured parameters:

\begin{equation}\label{eq:emit_0}
\varepsilon^2_0 = \varepsilon_m^2 - \varepsilon_m \left (\beta_m \frac{d_1^2 + d_2^2}{3L^2} + \frac{1+\alpha_m^2}{\beta_m} \frac{d_1^2}{3} - \alpha_m \frac{2d_1^2}{3L} \right ) + \frac{d_1^2 d_2^2}{9L^2} \quad .
\end{equation}
\\

The error on the measured emittance due to a finite slit size is therefore

\begin{equation}\label{eq:error}
\frac{\varepsilon_m - \varepsilon_0}{\varepsilon_0} = \left (1 - \frac{1}{\varepsilon_m} \left ( \beta_m \frac{d_1^2 + d_2^2}{3L^2} + \frac{1+\alpha_m^2}{\beta_m} \frac{d_1^2}{3} - \alpha_m \frac{2d_1^2}{3L} \right ) + \frac{1}{\varepsilon^2_m} \frac{d_1^2 d_2^2}{9L^2} \right ) ^{-\frac{1}{2}} - 1 \quad .
\end{equation}
\\


\section{Discussion}
The results in \cite{sander} differ from those derived in this paper by typos and a numerical coefficient in the expressions for $\alpha_m$ and $\varepsilon_m$, as well as by the absence of the last term in Eq. \ref{eq:emit_m}. Note that this term does not appear if the derivation is made by integration of the distribution in Eq. \ref{eq:dist}, expanded near the location of the slits, as it requires slit sizes much smaller than the width of both beam and beamlet.

As a numerical example of the effect, the error in measured emittance estimated with Eq. \ref{eq:error} using dimensions of the PXIE Allison scanner ($2d_1$ = 0.2~mm, $2d_2$ = 0.5~mm, and $L$ = 118~mm), with typical beam parameters at the end of the PXIE LEBT ($\alpha_m$ = -0.56 rad, $\beta_m$ = 0.33 m, and $\varepsilon_m$ = 14.4 mm mrad), is 3.2$\%$.


\section{Acknowledgements}
Fermilab is operated by Fermi Research Alliance, LLC, under contract DE-AC02-07CH11359 with the United States Department of Energy.

\appendix
\section{Exponential Integration Identities}

\begin{equation}\label{eq:eu}
\int^{\infty}_{-\infty} e^{-ax^2 - 2bx} dx = \sqrt{\frac{\pi}{a}} e^{\frac{b^2}{a}} \quad , \enspace \rm{where} \enspace a>0 \quad .
\end{equation}

\begin{equation}\label{eq:ueu}
\int^{\infty}_{-\infty} x e^{-ax^2 + bx} dx = \frac{\sqrt{\pi}b}{2a^{3/2}} e^{\frac{b^2}{4a}} \quad , \enspace \rm{where} \enspace Re(a)>0 \quad .
\end{equation}

\begin{equation}\label{eq:u2eu}
\int^{\infty}_{-\infty} x^2 e^{-ax^2 - bx} dx = \frac{\sqrt{\pi}(2a+b^2)}{4a^{5/2}} e^{\frac{b^2}{4a}} \quad , \enspace \rm{where} \enspace Re(a)>0 \quad .
\end{equation}

\end{document}